\documentclass[prb,twocolumn,superscriptaddress,showpacs,letterpaper]{revtex4}
\usepackage{graphicx}
\begin{document}
\newtheorem{theorem}{Theorem}
\newtheorem{acknowledgement}[theorem]{Acknowledgement}
\newtheorem{algorithm}[theorem]{Algorithm}
\newtheorem{axiom}[theorem]{Axiom}
\newtheorem{claim}[theorem]{Claim}
\newtheorem{conclusion}[theorem]{Conclusion}
\newtheorem{condition}[theorem]{Condition}
\newtheorem{conjecture}[theorem]{Conjecture}
\newtheorem{corollary}[theorem]{Corollary}
\newtheorem{criterion}[theorem]{Criterion}
\newtheorem{definition}[theorem]{Definition}
\newtheorem{example}[theorem]{Example}
\newtheorem{exercise}[theorem]{Exercise}
\newtheorem{lemma}[theorem]{Lemma}
\newtheorem{notation}[theorem]{Notation}
\newtheorem{problem}[theorem]{Problem}
\newtheorem{proposition}[theorem]{Proposition}
\newtheorem{remark}[theorem]{Remark}
\newtheorem{solution}[theorem]{Solution}
\newtheorem{summary}[theorem]{Summary}   
\def\r{{\bf{r}}}
\def\i{{\bf{i}}}
\def\j{{\bf{j}}}
\def\m{{\bf{m}}}
\def\k{{\bf{k}}}
\def\kt{{\tilde{\k}}}
\def\mt{{\hat{t}}}
\def\mG{{\hat{G}}}
\def\mg{{\hat{g}}}
\def\mGa{{\hat{\Gamma}}}
\def\mS{{\hat{\Sigma}}}
\def\mT{{\hat{T}}}
\def\K{{\bf{K}}}
\def\P{{\bf{P}}}
\def\q{{\bf{q}}}
\def\Q{{\bf{Q}}}
\def\p{{\bf{p}}}
\def\x{{\bf{x}}}
\def\X{{\bf{X}}}
\def\Y{{\bf{Y}}}
\def\F{{\bf{F}}}
\def\G{{\bf{G}}}
\def\bG{{\bar{G}}}
\def\mbG{{\hat{\bar{G}}}}
\def\M{{\bf{M}}}
\def\V{\cal V}
\def\tchi{\tilde{\chi}}
\def\tx{\tilde{\bf{x}}}
\def\tk{\tilde{\bf{k}}}
\def\tK{\tilde{\bf{K}}}
\def\tq{\tilde{\bf{q}}}
\def\tQ{\tilde{\bf{Q}}}
\def\si{\sigma}
\def\ep{\epsilon}
\def\hep{{\hat{\epsilon}}}
\def\al{\alpha}
\def\be{\beta}
\def\ep{\epsilon}
\def\bep{\bar{\epsilon}_\K}
\def\up{\uparrow}
\def\de{\delta}
\def\De{\Delta}
\def\up{\uparrow}
\def\dwn{\downarrow}
\def\ksi{\xi}
\def\etha{\eta}
\def\product{\prod}
\def\goto{\rightarrow}
\def\switch{\leftrightarrow}

\title{Dynamical mean field study of the Mott transition in the half-filled Hubbard model on a triangular lattice} 

\author{K.~Aryanpour}
\affiliation{Department of Physics, University of California, 
Davis, California 95616} 

\author{W.~E.~Pickett}
\affiliation{Department of Physics, University of California, 
Davis, California 95616}

\author{R.~T.~Scalettar} 
\affiliation{Department of Physics, University of California, 
Davis, California 95616}

\date{\today}

\begin{abstract}
We employ dynamical mean field theory (DMFT) with a Quantum Monte Carlo
(QMC) atomic solver to investigate the finite temperature Mott transition 
in the Hubbard model with the nearest neighbor hopping on a triangular lattice
at half-filling.  We estimate the value of the critical interaction to be $U_c=12.0 \pm 0.5$ in units of the hopping amplitude $t$
through the evolution of the magnetic moment, spectral function,
 internal energy and specific heat as the interaction $U$ and temperature
$T$ are varied. This work also presents a comparison between DMFT and
finite size determinant Quantum Monte Carlo (DQMC) and a discussion of
the advantages and limitations of both methods. 

\end{abstract}

\maketitle 

\section{Motivation}
\label{sec:introduction}            

Systems with triangular lattice structure have been a source of
attention mostly due to the frustration effects resulting from their
non-bipartite structure. As a result of the competition between the
frustration and strong electron correlations, these systems
exhibit a wide range of exotic phases. Recent studies of the
metal-insulator transition, superconductivity, and antiferromagnetism in
the organic compounds $\kappa$-(BEDT-TTF)$_{2}$X with X as an anion
\cite{kanoda,mckenzie}, discovery of superconductivity in
Na$_{x}$CoO$_{2}$.yH$_{2}$O,\cite{takada} and the recent discovery of
the Mott transitions in 0.33 monolayers of Sn on Ge(111) at $30$
K,\cite{cortes} as a few examples, underline the importance of these
systems and their physics.   

Theoretical work has been dedicated to the discovery of the magnetically ordered phases in the ground state of the Hubbard and $t-J$ models on triangular lattices as a function of the on-site electron-electron Coulomb interaction $U$ at 
different fillings.\cite{krishnamurthy1,krishnamurthy2,capone,morita,srivastava,weber}
 Unlike the square lattice at half-filling, which is a Mott insulator with 
antiferromagnetic order at arbitrarily small values of $U/t$, the ground state 
of a triangular lattice has a variety of magnetically ordered and disordered 
phases. This is due to the lack of perfect nesting in the non-interacting 
Fermi surface of a triangular lattice at half-filling. 

For triangular lattices, according to the Hartree-Fock calculations of
Krishnamurthy and co-workers, \cite{krishnamurthy1,krishnamurthy2} the
Mott transition occurs from a paramagnetic metal to a paramagnetic
insulator at half-filling for values of $U$ larger than the band width
$W=9t$.  A variety of the physical properties of the triangular lattices
including tendencies towards superconductivity within the small to
intermediate $U$ regime have been studied at finite temperature using
correlated electron approaches which go beyond mean field theory, such
as the fluctuation exchange approximation (FLEX)\cite{renner} or
one-loop renormalization-group.\cite{honerkamp} However, when
 $U/W\geq1$, a more powerful cluster solving
technique such as the Quantum Monte Carlo (QMC) is required to accurately 
describe the phase transition. Finite size
lattice determinant Quantum Monte Carlo (DQMC) has already been employed
for triangular and kagom\'e lattices.\cite{bulut,refolio} Unfortunately,
the finite size lattice QMC method incurs sign problems at both low
temperatures and away from half-filling. In addition, due to the finite
size nature of the problem, the non-local correlations are always
overestimated and the system consisting of a finite size cluster becomes an 
insulator as soon as the correlation length reaches the size of the lattice 
at low enough $T$. Thus, the gaps or pseudogaps are also overestimated and 
are not guaranteed to persist as one reaches the thermodynamic
limit.\cite{moukouri} However, finite size effects can be reduced by size 
scaling and in general become less important as $U$ increases.

In this work, we employ the dynamical mean-field theory
(DMFT)\cite{metzner,muller,georges} to study the Mott transition for the
Hubbard model on a triangular lattice at half-filling.  DMFT describes systems in the thermodynamic limit and at the same time suppresses the physics due to the non-local correlations through coarse-graining over all momentum degrees of freedom. Therefore, susceptibilities can diverge, as in the thermodynamic limit but the Mott gap in the strong-coupling regime is always smaller than its
counterpart given by a finite size lattice approach. The combination of
the DMFT and QMC also does not incur sign problems at either low
temperatures or any fillings which makes the low temperature physics
more accessible and reliable. 

DMFT is a more suitable approximation for the triangular lattice systems
as opposed to those with square lattice structure. For instance, DMFT is
incapable of describing the metal-insulator transition on the square
lattice in the weak coupling regime. This is a consequence of the
suppression of the short-range antiferromagnetic correlations
responsible for this transition at small $U$ by the single site nature
of DMFT.\cite{moukouri} However, in triangular lattices, due to the
frustration of the lattice structure, even in the presence of non-local
correlations, the short-range antiferromagnetic correlations are greatly
suppressed. As a result, heavy quasiparticles form at the Fermi energy
reminiscent of the DMFT physics.\cite{imai} 

DMFT has already been employed for studying triangular lattices using
the exact diagonalization and Lanczos techniques.\cite{merino} In this
paper, we present for the first time, the combination of DMFT and QMC
for solving the Hubbard model at half-filling on a triangular lattice.
By computing the density of states at different values of interaction
$U$ and temperature $T$, we estimate the value of the critical interaction 
for the Mott transition to be $U_c/t=12.0\pm 0.5$, less than the predicted
value of $15$ in Ref.~\onlinecite{merino}, but, as expected, larger than
the mean field result.  In addition, the Mott transition
features are also manifest in the evolution of the magnetic moment,
total internal energy and the specific heat as functions of
on-site interaction $U$ and temperature $T$.  

\section{Model and Formalism}
\label{sec:formalism}

We consider the repulsive Hubbard model on a triangular lattice,

\begin{eqnarray}
\label{eq:rep-hub-mod}
H=&-&t\sum_{<ij>,\sigma}(c^{\dag}_{i\sigma}c_{j\sigma} +
c^{\dag}_{j\sigma}c_{i\sigma})
\nonumber \\
&-& \mu\sum_{i\sigma}c^{\dag}_{i\sigma}c_{i\sigma}+U\sum_{i}
n_{i\uparrow}n_{i\downarrow}\,,
\end{eqnarray}
with $t$ the hopping amplitude, $\mu$ the chemical potential and $U$ the
on-site Coulomb interaction between the electrons of opposite spins
residing on lattice site $i$. Note that unlike the square lattice, on a
triangular lattice, due to the lack of particle-hole symmetry, $\mu=U/2$
does not yield half-filling. Also, this model exhibits quite
different physics under the change of the sign of $t$, especially away from
half-filling. The dispersion in the $U=0$ limit including the nearest
neighbor hopping only is, 
\begin{eqnarray}
\label{eq:trig-disp}
\epsilon_{k}=-2t[\cos(k_{x})+\cos(\frac{k_{x}}{2}+\frac{\sqrt{3}}{2}k_{y})+
\nonumber \\ \cos(\frac{k_{x}}{2}-\frac{\sqrt{3}}{2}k_{y})]\,.
\end{eqnarray}
The resulting noninteracting density of states is markedly
asymmetric about $E=0$, extending from $E=-6t$ to 
$E=+3t$, with a van Hove singularity at $E=t$.  

The DMFT approach to tight binding Hamiltonians starts by coarse-graining
the Green's function,
\begin{eqnarray}
\label{eq:Green-CG}
G_{\sigma}(i\omega_{n})=\frac{1}{N}\sum_{k}G_{\sigma}(k,i\omega_{n})=\nonumber \\ \frac{1}{N}\sum_{k}\frac{1}{i\omega_{n}-\epsilon_{k}+\mu-\Sigma_{\sigma}(i\omega_{n})}\,,
\end{eqnarray}
where $\omega_n=(2n+1)\pi T$ is the Matsubara fermionic frequency and
$\Sigma_{\sigma}(i\omega_{n})$ the local self-energy for spin $\sigma$.
Note that the self-energy $\Sigma_{\sigma}(i\omega_{n})$ in DMFT is
local (momentum independent).  In the first DMFT cycle iteration, we
chose $\Sigma_{\sigma}(i\omega_{n})=0$.  The bath Green's function is
then computed,
\begin{equation}
\label{eq:bath-Gf}
{\cal G}_{\sigma}(i\omega_{n})= (G_{\sigma}^{-1}(i\omega_{n})+\Sigma_{\sigma}(i\omega_{n}))^{-1}\,.
\end{equation}
${\cal G}_{\sigma}(i\omega_{n})$ is used as an input
to the QMC solver. Our DMFT-QMC solver is based on the Hirsch-Fye
algorithm \cite{hirsch-fye1,hirsch-fye2} in which the interacting
two-body term in the Hubbard model is decomposed into two one-body
density terms coupled to a Hubbard-Stratonovich field. Through the
stochastic averaging over the most probable configurations
of these Ising like fields, one calculates the fully interacting Green's
function on a single site of a lattice $G_{\sigma}(i\omega_{n})$.
The local self-energy $\Sigma_{\sigma}(i\omega_{n})$ is then calculated
from the full and bath Green's functions $G_{\sigma}(i\omega_{n})$ and
${\cal G}_{\sigma}(i\omega_{n})$ respectively,   
\begin{equation}
\label{eq:self-energy}
\Sigma_{\sigma}(i\omega_{n})= {\cal G}_{\sigma}^{-1}(i\omega_{n})-G_{\sigma}^{-1}(i\omega_{n})\,.
\end{equation}
This self-energy is again inserted in Eq.\ref{eq:Green-CG}, and
the self-consistent process continues until convergence.

The final product of the DMFT-QMC cycle is the full Green's function in
imaginary time $G(\tau)$ which is used to compute the density
of states (DOS) $N(\omega)$,

\begin{equation}
\label{eq:Gtau}
G(\tau)= T\sum_{n}e^{-i\omega_n\tau}G(i\omega_n)= \int d\omega\frac{N(\omega)}{1+e^{-\beta\omega}}e^{-\omega\tau}\,.
\end{equation}
$N(\omega)$ must be calculated by inverting
Eq.\ref{eq:Gtau}, which is an ill-posed problem due to the statistical
errors in $G(\tau)$ from the QMC simulations. This task in this paper is
done by using the maximum entropy (MaxEnt) analytical continuation
technique developed by Jarrell and Gubernatis.\cite{jarrell1} This
technique is based on Bayesian inference in which an entropy function
with respect to an original default model is defined as a function of
$N(\omega)$. The best $N(\omega)$  is the one that maximizes this
entropy function for a given $G(\tau)$.  

We also present results for the double occupation,
\begin{equation}
\label{eq:dbleoccp}
<D>= <n_{i\uparrow}n_{i\downarrow}>\,,
\end{equation}
which is suppressed by the on-site repulsion.
The double occupation is related to the magnetic moment, 
\begin{eqnarray}
\label{eq:MagMom}
<m_{zi}^{2}>=<(n_{i\uparrow}-n_{i\downarrow})^2> 
= \nonumber \\ <n_{i\uparrow}>+<n_{i\downarrow}> -2<D>\,.
\end{eqnarray}

The total internal energy can also be computed in this formalism in
terms of the full Green's function and double occupation:
\begin{equation}
\label{eq:int-eng}
E_{int}=\frac{T}{N}\sum_{nk\sigma}[\epsilon_kG_{\sigma}(k,i\omega_n)]e^{i\omega_n0^{+}}+UD\,,
\end{equation}
with $D$ defined as in Eq.\ref{eq:dbleoccp}. Eq.\ref{eq:int-eng} is
equivalent to the Galitskii-Migdal expression for the total internal
energy.\cite{galitskii} The specific heat at constant volume is 
\begin{equation}
\label{eq:spec-heat}
C_{V}=\big(\frac{\partial E_{int}}{\partial T}\big)_{V}\,.
\end{equation}
As seen later in this article, the variation of the total internal
energy and specific heat as a function of $U$ and $T$ will exhibit 
features reflecting the Mott metal-insulator phase transition. 
 
\section{Results}
\label{sec:results}

Before presenting our data for the density of states, we comment
on some of the details of the maximum entropy procedure, since
careful characterization of the data is essential to obtain reliable
results.
In this work, the DMFT-QMC cycle as described in section
\ref{sec:formalism} converges at a desired tolerance which is about few
percent for the time Green's function $G(\tau)$ . The chemical potential
is also adjusted simultaneously to keep the total occupation at
half-filling. For the QMC runs, we adopt the criterion
$U(\Delta\tau)^2<0.05$ with $\Delta\tau=\beta/L$ for finding $L$, the
number of the time slices. $L$ can be small at high temperatures,
however, even for temperatures as high as the bandwidth, we choose the
lower limit of $L=40$. The number of the lattice sweeps for the QMC runs
varies between $N=2\times 10^5$ to $1\times 10^6$ depending on the
values of $U$ and $T$ and the reliability of the MaxEnt results. The
outputs of the cycle are the fully interacting time Green's function,
the magnetic moment on a single site and total internal energy for the
temperature and interaction at which the QMC cycle has been running. The
full $G(\tau)$ is employed to calculate the DOS through Eq.\ref{eq:Gtau}
using the MaxEnt technique. \cite{jarrell1}

Care must be taken when using MaxEnt as the results can strongly depend
on the quality of the QMC data. MaxEnt, as described in
Ref.~\onlinecite{jarrell1}, requires statistically independent and Gaussianly
distributed measurements of $G(\tau)$ for every time slice $\tau$ from
QMC. Both these two properties are often lacking in the 
$G(\tau)$ data initially obtained from the QMC output because
of the presence of correlations between the measurements.   The solution
is a careful rebinning of the $G(\tau)$ data which eliminates these
correlations and yields Gaussianly distributed data.
Correlations between different time slices are also eliminated in the
MaxEnt code by diagonalizing a covariance matrix coupling different time
slices together.\cite{jarrell1} 

The other concern is a proper choice for the
MaxEnt default model. In general, for good QMC data, the choice of the
default model should not change the qualitative features of the spectral
function significantly. Among the available default models, the
Gaussian and second order perturbation models have been most commonly
utilized. In our work, we adopt the latter and perform the so called
annealing method in which we start with the second-order perturbation
theory default model at a high temperature close to the band width. We
then use the MaxEnt output for that temperature as the default model for
a slightly lower temperature and continue until we arrive at the lowest
desired $T$. Finally, as introduced in Ref.~\onlinecite{jarrell1}, there are
different methods of doing the MaxEnt including the historic, classic
and Bryan. We utilized both classic and Bryan techniques in our
calculations and found that both these two methods give qualitatively
similar results.        

\begin{figure}
\includegraphics[width=3.2in]{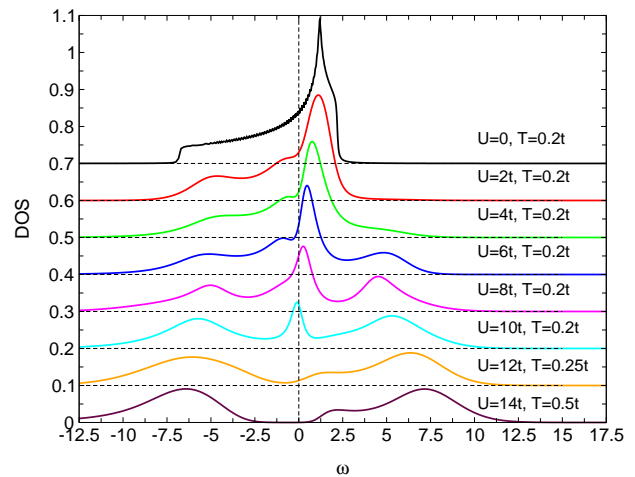}
\caption[a]{(color on-line) The evolution of the density of states at half-filling for the triangular lattice as a function of $U$. For clarity, the base lines have been shifted in steps of $0.1$ for different values of $U$. The $\omega=0$ line determines the location of the Fermi energy.
Three behaviors are visible:  At weak coupling the density of states
has a finite value at $\omega=0$.  At large coupling the density of
states vanishes, signaling the opening of the Mott insulating gap.
At intermediate coupling the density of states exhibits a Kondo 
resonance at $\omega=0$.
}
\label{dos.evolve}
\end{figure} 

In Fig.\ref{dos.evolve}, we present the evolution of the DOS as a
function of $U$ at low temperatures.  At $U=0$, the DOS shows a Van Hove
singularity near $\omega = t$. As the interaction is gradually turned
on, there is an overall broadening of the density of states, including a
smearing out of the Van Hove singularity. The DOS at $U=8t$ and $T=0.2t$
exhibits the onset of the Abrikosov-Suhl (Kondo) resonance in which
there appears a sharp quasi-particle peak near the Fermi energy with two shoulders, the upper and lower Hubbard bands, around
it. For $U=10t$ and $T=0.2t$, the peak lines up with the Fermi energy and shoulders are approximately located at $\pm 5t \approx \pm U/2$.
At $U=12t$, and $T=0.25t$,  we clearly observe a different trend in which
the Kondo resonance is absent and the DOS has a pronounced minimum at
the Fermi surface, indicating that the system has been driven to the
metal-insulator transition regime. For $U=14t$, and $T=0.5t$, the (Mott) 
gap is well established and the system becomes an insulator. 

\begin{figure}
\includegraphics[width=2.8in]{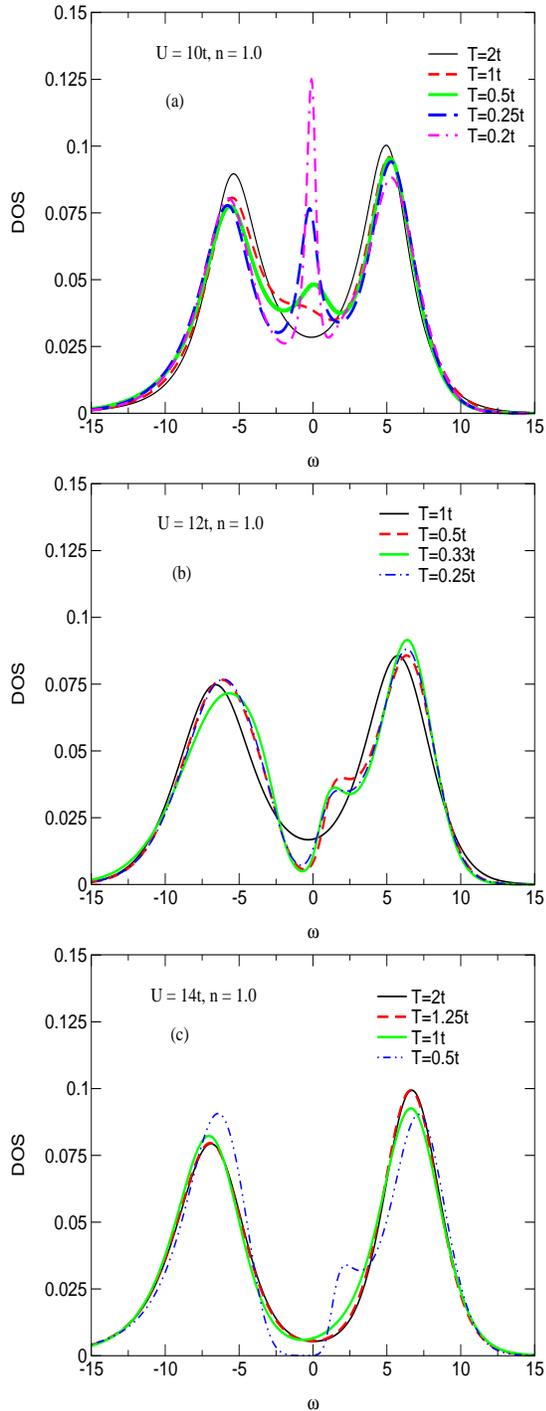}
\caption[a]{(color on-line) Panel (a): the DOS for $U=10t$ at half-filling and different temperatures.  As the temperature is lowered a sharp Kondo resonance develops at $\omega=0$.  The system is metallic.  Panel (b): the same results as for panel (a) for $U=12t$.  The system is on the verge of insulating behavior. Panel (c): $U=14t$. A Mott gap has opened at $T=0.5t$.} 
\label{dos.compare}
\end{figure}

\begin{figure}
\includegraphics[width=3.4in]{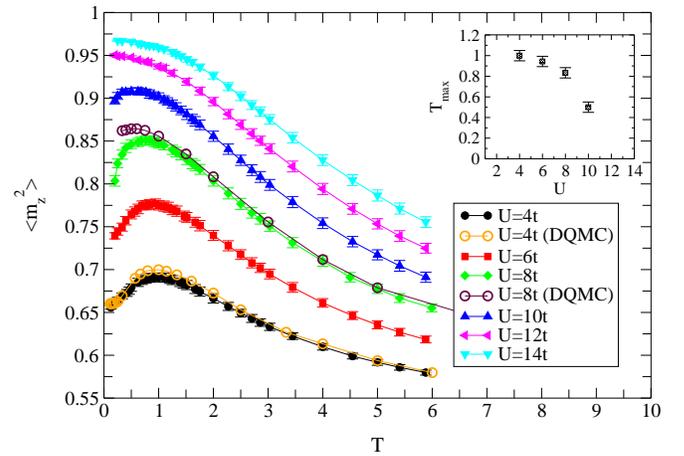}
\caption[a]{(color on-line) Variation of the magnetic moment with temperature for different values of interaction $U$ at half-filling. $<m_z^2>$ drops down close to its minimum $0.50$ at large temperatures (cut off in this figure). The comparison between the DQMC and DMFT at $U=4t$ and $8t$ is also presented. The inset shows how $T_{max}$, the temperature for the maximum $<m_z^2>$,  drops as $U$
increases.}
\label{Mag-Mom-T}
\end{figure}

To better address the role played by temperature in the metal-insulator
transition, in Fig.\ref{dos.compare} the DOS for three values of
$U=10t$, $12t$ and $14t$ and different temperatures have been plotted. At
$U=10t$, as $T$ decreases, the DOS evolves towards a sharp
Abrikosov-Suhl resonant quasi-particle peak.  However, $U=12t$
exhibits a completely different trend. As the temperature drops, the DOS
evolves towards the opening of a gap. At $T=0.25t$, we noticed that the
MaxEnt results are of less reliability due to the correlations in the
QMC data which could not be totally removed up to $10^6$ QMC lattice
sweeps by rebinning the data.  (It is, of course, typical of Monte Carlo
simulations that statistical fluctuations are largest near phase
transitions) This is additional evidence that $U=12t$ is close to the critical value for the opening of a gap but we were not able to determine conclusively whether the gap exists at $U=12t$ if one can reach low enough temperatures. On
the other hand, by the time $U=14t$ (at $T=0.5t$), the MaxEnt results
show a pronounced Mott gap in the DOS. Therefore, it is clear that the
critical value of $U_c$ must lie very close to $U=12t$, leading to our estimate $U=12t\pm0.5t$. This value is smaller than the value $U_{c}\approx15t$ obtained in Ref.~\onlinecite{merino} using the DMFT with exact diagonalization and Lanczos techniques as the impurity solvers, and comparable to $U_{c}\approx12t$ in Ref.~\onlinecite{capone} with exact diagonalization on $12$ site lattices. At both values of $U=12t$ and $14t$,  we also observe the appearance of a feature on
the right hand side of $\omega=0$. This feature is likely a remnant of
the peak in the non-interacting density of states-- it disappears as $U$
is increased to larger values and is completely absent for $U=16t$. 

Another signature of the metal-insulator (Mott) transition is manifest in
the behavior of the magnetic moment defined in Eq.\ref{eq:MagMom} as a
function of temperature $T$ and interaction $U$. As shown in
Fig.\ref{Mag-Mom-T}, for small to intermediate values of $U$, $<m_z^2>$
reaches a maximum at a value of $T_{max}$ which is characteristic of a
Fermi-liquid dominated by spin fluctuations at low
temperatures.\cite{georges2,georges3,moukouri} At low temperatures with
an entropy of $\gamma T$ per particle, a Fermi liquid gains free energy
upon heating by trying to localize the electrons in order to take
advantage of a larger spin entropy.\cite{georges2,georges3} The inset in
Fig.\ref{Mag-Mom-T} shows how $T_{max}$ decreases with $U$. However,
at the onset of the Mott transition, $<m_z^2>$ becomes a monotonically
decreasing function of $T$ as seen in Fig.~\ref{Mag-Mom-T} for
$U=12t$ and $14t$.  In a localized Mott phase, the spin
entropy at $T\approx0$ is finite and the system cannot gain free energy
by further localizing the electrons upon heating. 
Comparison has also been made
between DQMC on a $6\times 6$ finite size lattice and DMFT in
Fig.\ref{Mag-Mom-T} for values of $U=4t$ and $8t$. The agreement between the
two approaches is very close at $U=4t$ down to $T=0.125t$.
However, for $U=8t$, at low temperatures, the DQMC approach exhibits a
rather flat region for $<m_z^2>$ below $T\approx 1t$ in place of the
downward curve in DMFT. This flat feature is similar to what is observed
at larger values of $U$ in DMFT as a precursor to the insulating phase.
This is typical of any finite size lattice approach such as DQMC in
which at low enough temperatures, once the correlation length exceeds
the size of the lattice, the system turns insulating. In the DMFT, on
the other hand, the system is in the thermodynamic limit and
therefore, the insulating phase features do not appear except at large
enough interaction $U$ values. Nevertheless, DQMC has the advantage of
including the non-local correlations which are ignored in DMFT. These
correlations work in favor of the gap formation. Thus, DMFT 
overestimates the value of $U$ for the formation of the gap in the Mott
transition.      

\begin{figure}
\includegraphics[width=3.0in]{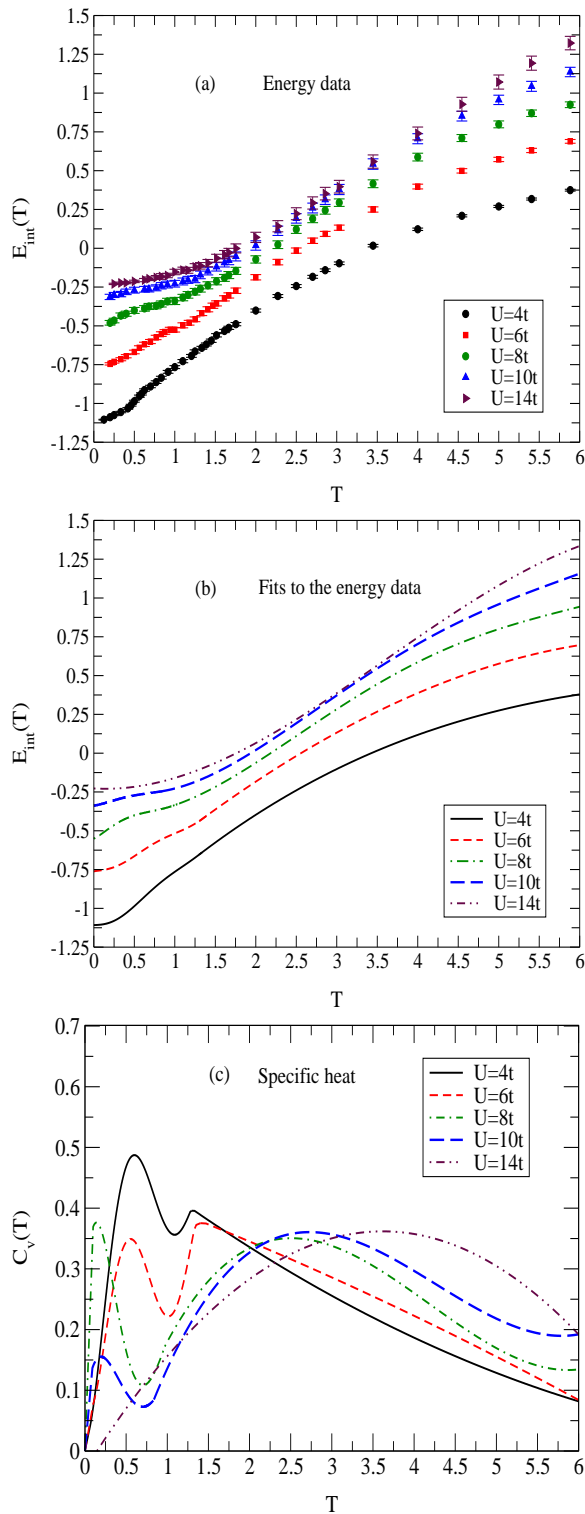}
\caption[a]{(color on-line) Panel (a): Variation of the total internal energy versus temperature $T$ for different values of interaction $U$ including the error bars. Panel (b): Polynomial fits to data in panel (a). Panel (c): The specific heat as a function of temperature $T$ for different values of interaction $U$ taken by analytical differentiation of polynomials in
panel (b). }
\label{thermodynamics}
\end{figure}

The evolution of the total internal energy and specific heat defined in
Eq.\ref{eq:int-eng} and Eq.\ref{eq:spec-heat} respectively also exhibits
signatures of a metal-insulator phase transition at large interaction
$U$ values and low temperatures $T$. In Fig.\ref{thermodynamics}, panel
(a), the total internal energy has been plotted as a function of
temperature $T$ for different values of interaction $U$. We have
employed a polynomial fitting procedure developed by
Duffy and Moreo \cite{duffy-moreo} and obtained the curves
presented in panel (b). By analytically differentiating the polynomial
fits with respect to temperature, we arrive at results for the specific
heat plotted in panel (c). For values of $U=4t$ to $10t$, we observe a
two-peak structure in the $C_V$ consistent with the results in
Ref.~\onlinecite{georges3,duffy-moreo} for the infinite dimensional and $2D$
hypercubic Hubbard models respectively. The peak at low
temperatures corresponds to the local spin fluctuations analogous to the
formation of the Kondo singlets in the Kondo model. It is also a signature for
the validity of the Fermi liquid picture as it indicates the existence of a conducting electron bath around the magnetic impurity.\cite{georges3} The upper peak at high temperatures, however, corresponds to on-site charge
fluctuations. Separation between these two peaks occurs at $U=4t$ and develops 
up to $U=8t$ resulting in a sharp lower peak. 

As prescribed in Ref.~\onlinecite{duffy-moreo}, the lower and higher temperature segments of $E_{int}(T)$ have been fit with two different polynomials which meet around $T\approx1t$. This prescription explains the rather sharp feature in the $C_V$ curve at $U=4t$ in Fig.\ref{thermodynamics}, panel (c), around $T\approx1.25t$. However, the
necessity of using two polynomials originates in the existence of
two separate temperature regions, one dominated by spin and the other
by charge fluctuations. The lower peak dwindles at $U=10t$ as a sign of approaching the insulating phase.\cite{paiva} It is also striking that, similar to the result of Ref.~\onlinecite{georges3}, within the Fermi liquid range up to $U=10t$, all $C_V$ curves more or less intersect at the same temperature close to
$T=2.1t$. The picture however is different at $U=14t$ when the
insulating phase sets in. The lower peak completely vanishes and the
$C_V$ curve goes to zero at a finite value of $T$. This is consistent
with the behavior of $C_V$ in the presence of a gap $\Delta$
going as $e^{-\Delta/T}$ at low temperatures, and
having a broad peak at high temperatures resulting from charge
fluctuations in the upper Hubbard band.\cite{georges3} For $U=12t$,
$E_{int}$ results were too noisy to be conclusive and therefore are not
presented. This was apparently due to $U=12t$ being close to the critical 
$U_c$ which made it difficult to achieve convergence in the simulation.
 
\begin{figure}
\includegraphics[width=3.1in]{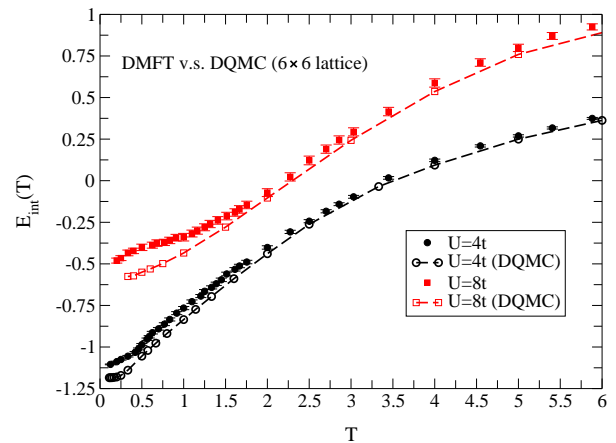}
\caption[a]{(color on-line) Comparison between DMFT and DQMC on a $6\times6$ finite size
lattice for the total internal energy versus temperature $T$ for
interaction values of $U=4t$ and $8t$.}
\label{Eint-T-comp}
\end{figure}

Lastly, Fig.\ref{Eint-T-comp} presents a comparison between DMFT and
DQMC for $E_{int}(T)$ at $U=4t$ and $8t$. Similar to the trend observed
for $<m_z^2>$ in Fig.\ref{Mag-Mom-T}, at high enough $T$, the agreement
betweeen the two approaches is clearly manifest in the results. For
$U=4t$ at low $T$, DQMC exhibits an insulating phase feature in the
rather flat tail of the $E_{int}(t)$ curve near $T=0$, typical of a finite
size approach as discussed earlier. For $U=8t$, however, the effect of
non-local correlations becomes more important and at the same time,
finite size effects diminish. Hence, DQMC features may be more realistic
in showing insulating phase signatures. Nevertheless, $U=8t$ is still
below the bandwidth $W=9t$ and we do not expect to have entered the
insulating phase as yet. Unfortunately, mostly due to the sign problem
in DQMC, attempts in going to lower temperatures and higher $U$
values have not yet been successful. Thus, it is not clear to us which
approach more plausibly describes the physics for the intermediate value
of $U=8t$.   
  
In order to bring the DQMC and DMFT treatments into full agreement,
it would be necessary to extrapolate both to a zero value of  the 
discretization size of the inverse temperature $\beta$ and also
to extrapolate DQMC and DMFT to infinite spatial cluster size and
full momentum resolution (the dynamical cluster approximation (DCA)),
respectively.

\section{Conclusions}
\label{sec:conclusions}

In summary, we have performed the first implementation of the DMFT using
QMC as the solver for the half-filled triangular lattice Hubbard model.
Our approach confirms the previous predictions that for large enough
$U/W$, there is a metal-insulator (Mott) phase transition from
a paramagnetic metal to a paramagnetic insulator at finite temperature.
We demonstrate this phase transition by presenting the behavior of the
DOS, magnetic moment on a single site, total internal energy and
specific heat as functions of the interaction $U$ and temperature $T$.
Our results suggest a critical value of interaction $U_{c}/t=12.0\pm0.5$.  This is consistent with the value $U_{c}\approx12t$ from exact
diagonalization on $12$ site lattices \cite{capone} and
$U_{c}\approx15t$ from DMFT with exact diagonalization and Lanczos
impurity solvers.\cite{merino}. We have also investigated the
intermediate interacting regime which is characterized by the presence
of an Abrikosov-Suhl (Kondo) resonance. 

DMFT is a more suitable approximation for studying the Mott transition
on a triangular lattice compared to a square lattice. The reason is that
in a triangular lattice, due to the frustration of the lattice
structure, the short-range antiferromagnetic correlations are suppressed
and do not play a major role in the metal-insulator transition at small
$U$.\cite{imai} Hence, neglecting these correlations, as done by DMFT,
should still give us a qualitatively correct physical picture at least
in the large $U$ limit. The combination of
the DMFT and QMC also does not encounter the difficulties related to the sign
problem at low temperatures and away from half-filling and also finite
size effects associated with DQMC at low $T$. Thus, the low temperature
physics is better described by the DMFT due to being in the
thermodynamic limit. Nevertheless, DQMC has the advantage of
incorporating non-local correlations which act in favor of gap formation
and DMFT overestimates the value of $U_c$ because of lacking these
correlations. 

Due to the single site nature of the DMFT, we were not able to address
the physics of magnetically ordered phases for triangular lattices as
one approaches the $T=0$ limit. These studies have already been
performed using the DQMC techniques.\cite{bulut,refolio} However, the
combination of the dynamical cluster approximation (DCA) and QMC
\cite{hettler1,hettler2,jarrell2} can also be considered a promising
future candidate for these studies as a complementary approach to DQMC.
Work on different magnetically ordered phases at different coupling
regimes, including ferromagnetism away from half-filling as discussed in Ref.\onlinecite{merino} is in progress.    

\section{Acknowledgments}
\label{sec:acknowledgments}

We acknowledge useful conversations with M.~Jarrell, A.~Macridin, and
S.~Savrasov. This
research was supported by DOE DE-FG03-03NA00071. 
and DE-FG01-06NA26204.

\end{document}